\documentclass[11pt]{article}
\usepackage[utf8]{inputenc}
\usepackage[T1]{fontenc}
\usepackage{hyperref}
\usepackage{amsmath,amssymb}
\usepackage[active]{srcltx}
\usepackage[all]{xy}
\input diagxy
\usepackage{xcolor}
\font\fr=eufm10 scaled \magstep 1 
\newtheorem{teor}{Theorem}
\newtheorem{prop}{Proposition}

\newtheorem{definition}{Definition}

\newtheorem{remark}{Remark}

\def\beq{\begin{equation}}
\def\eeq{\end{equation}}
\def\bea{\begin{eqnarray}}
\def\eea{\end{eqnarray}}
\def\beann{\begin{eqnarray*}}
\def\eeann{\end{eqnarray*}}
\def\beasn{\begin{sneqnarray}}
\def\eeasn{\end{sneqnarray}}
\def\ben{\begin{enumerate}}
\def\een{\end{enumerate}}
\def\bit{\begin{itemize}}
\def\eit{\end{itemize}}
\def\proof{ (\emph{Proof\/}) }
\newcommand{\ds}{\displaystyle}
\def\derpar#1#2{\frac{\partial{#1}}{\partial{#2}}}

\def\qed{\ifvmode\Realemovelastskip\fi
{\unskip\nobreak\hfil\penalty50\hbox{}\nobreak\hfil \hbox{\vrule
height1.2ex width1.2ex}\parfillskip=0pt \finalhyphendemerits=0
\par\smallskip}}

\def\vf{\mbox{\fr X}}
\def\df{{\mit\Omega}}
\def\Lag{{\cal L}}
\def\L{{\cal L}}

\def\FL{{\cal FL}}

\def\d{{\rm d}}

\def\Real{\mathbb{R}}
\def\R{\mathbb{R}}

\def\Tan{{\rm T}}

\def\inn{\mathop{i}\nolimits}
\def\Cinfty{{\rm C}^\infty}

\title{\sc An overview of the Hamilton--Jacobi theory:
 the classical and geometrical approaches and some extensions and applications}
\author{\sffamily 
\sc 
Narciso Rom\'an-Roy
\thanks{narciso.roman@upc.edu\,({\it ORCID}:\,0000-0003-3663-9861).}
\\[1ex]
\normalsize\itshape\sffamily 
Dept. of Mathematics.
Universitat Polit\`ecnica de Catalunya.\\
\normalsize\itshape\sffamily 
Ed. C3, Campus Nord.
08034 Barcelona, Spain.}

\begin{document}

\maketitle

\pagestyle{myheadings}

\thispagestyle{empty}

\begin{abstract}
This work is devoted to review the modern geometric description of the 
Lagrangian and Hamiltonian formalisms of the Hamilton--Jacobi theory.
The relation with the ``classical'' Hamiltonian approach 
using canonical transformations is also analyzed.
Furthermore, a more general framework for the theory
is also briefly explained.
It is also shown how, from this generic framework, 
the Lagrangian and Hamiltonian cases of the theory for dynamical systems are recovered, 
and how the model can be extended to other types of physical systems, 
such as higher-order dynamical systems and (first-order) classical field theories
in their multisymplectic 
formulation.
\end{abstract}

\noindent
  {\bf Key words}: 
\textsl{Hamilton--Jacobi equations,  Lagrangian 
and Hamiltonian formalisms, higher--order systems, classical field theories,
symplectic and multisymplectic manifolds, fiber bundles.}

\bigskip

\vbox{\raggedleft AMS s.\,c.\,(2020): 
{\it Primary\/}: 35F21, 70H03, 70H05, 70H20. \\
{\it Secondary\/}: 53C15, 53C80, 70G45, 70H50, 70S05.}\null

\markright{{\rm N. Rom\'an-Roy},
    {\sl Hamilton--Jacobi theory: classical and geometrical approaches.}}


\medskip
\setcounter{tocdepth}{2}
{
\small
\def\addvspace#1{\vskip 1pt}
\parskip 0pt plus 0.1mm
\tableofcontents
}

\newpage


\section{Introduction}

The Hamilton--Jacobi theory is a topic of interest in mathematical physics
since it is a way to integrate systems of first-order ordinary differential equations
(Hamilton equations in the standard case).
The classical method in Hamiltonian mechanics consists in 
obtaining a suitable canonical transformation which leads 
the system to equilibrium \cite{Ar,JS,LL-mec,Sa71},
and is given by its generating function. This function is  
the solution to the so-called {\sl Hamilton--Jacobi equation},
which is a partial differential equation.
The {\sl ``classical''  Hamilton--Jacobi problem} consists in finding this canonical transformation.
Because of its interest, the method was generalized in other
kinds of physical systems; such as, for instance,
singular Lagrangian systems \cite{Gomis1} or higher-order dynamics \cite{art:Constantelos84},
and different types of solutions have been proposed and studied
\cite{CEL-84,CL-83}.

Nevertheless, in recent times, a lot of research has been done to understand 
the Hamilton--Jacobi equation from a more general geometric approach,
and some geometric descriptions to the theory were done in
\cite{Ab,BT-80,EMS-2004,LM-87,MMMcq}.
From a geometric way, the above mentioned canonical transformation
is associated with a foliation in the the phase space of the system
which is represented by the cotangent bundle $\Tan^*Q$
of a manifold (the configuration manifold $Q$).
This foliation has some characteristic geometric properties:
it is invariant by the dynamics, transversal to the fibers of the cotangent bundle,
and Lagrangian with respect to the
canonical symplectic structure of $\Tan^*Q$
(although this last property could be ignored in some particular situations).
The restriction of the dynamical vector field in $\Tan^*Q$ to
each leaf  $S_\lambda$ of this foliation 
projects onto another vector field $X_\lambda$ on $Q$,
and the integral curves of these vector fields are one-to-one related.
Hence, the complete set of dynamical trajectories are recovered from the integral curves
of the complete family $\{ X_\lambda\}$ of all these vector fields in the base.
These geometric considerations can be done in an analogous way 
in the Lagrangian formalism and hence this geometrical picture 
of the Hamilton--Jacobi theory can be also stated for this formalism.
The {\sl geometric Hamilton--Jacobi problem} consists in finding
this foliation and  these vector fields $\{ X_\lambda\}$.

Following these ideas, the Lagrangian and Hamiltonian versions
of the Hamilton--Jacobi theory, for autonomous and non-autonomous mechanical systems, 
was formulated in another geometrical way
 in \cite{art:Carinena_Gracia_Marmo_Martinez_Munoz_Roman06}. 
The foundations of this geometric generalization are similar 
to those given in \cite{Ab,KV-1993}.
Later, this framework has been used to develop the Hamilton--Jacobi theory
for many other kinds of systems in physics. 
For instance, other applications of the theory are to the case of
singular Lagrangian and Hamiltonian systems \cite{HJteam-K,LMV-12,LMV-13,LOS-12},
higher-order dynamical systems \cite{CLRP-13,CLRP-14},
holonomic and non-holonomic mechanics
\cite{BFS-14,HJTeam2,leones2,leones1,LOS-12,OFB-11,blo},
and control theory \cite{BLMMM-12,Wang2}.
The theory is also been extended 
for dynamical systems described using other geometric structures, 
such as Poisson manifolds \cite{LMV-13,GP-2016}, Lie algebroids \cite{BMMP-10,LS-12},
contact manifolds (which model dissipative systems) \cite{Ll-2020,GP-2020},
and other geometric applications and generalizations:
\cite{BLM-12,HJTeam3,Wang3}.
Furthermore, in \cite{BDM-12,OBL-11}
the geometric discretization of the Hamilton--Jacobi equation was analyzed.
Finally, the Hamilton--Jacobi theory 
is developed for the usual covariant formulations of first-order classical field theories,
the $k$-symplectic and $k$-cosymplectic \cite{LMMSV-12,DeLeon_Vilarino}
and the multisymplectic ones \cite{LMM-09,LPRV-2017},
for higher-order field theories \cite{Vi-10,art:Vitagliano12},
for the formulation in the {\sl Cauchy data space\/} \cite{CLMV-14},
and for partial differential equations in general \cite{Vi-11,Vi-14}.

This review paper is devoted, first of all, to present,
in Section \ref{firstsec}, the foundations
of this modern geometric formulation of the Hamilton--Jacobi theory,
starting from the most general problem and explaining how to derive the standard
Hamilton--Jacobi equation for the Hamiltonian and the Lagrangian formalisms of autonomous mechanics.
After this, the notion of complete solution allows us to establish the
relation with the ``classical'' Hamilton--Jacobi theory based on canonical transformations,
which is summarized  in Section \ref{clasic},
where this relation is also analyzed 
(this topic had been already discussed in \cite{Vi-14}).
We also present briefly a more general geometric framework
for the Hamilton--Jacobi theory which was stated in \cite{HJTeam3},
from which we can derive the majority of the applications
of the theory to other kinds of physical systems,
including the case of autonomous dynamical systems.
This is done in Section \ref{co}.
Finally, among all the extensions of the theory, 
we have selected two of them for reviewing:
the case of higher-order autonomous dynamical systems
(Lagrangian and Hamiltonian formalisms), 
which is a direct application of the above general framework,
and the generalization to the Lagrangian and Hamiltonian multisymplectic
formalisms of first-order classical field theories, which can also be interpreted as a
special case of the general framework.
Both of them are treated in Section \ref{others}.

Throughout the work it is considered that
the manifolds are real, smooth and second countable. 
In the same way, all the maps are assumed to be smooth. 
The summation convention for repeated cross indices is also assumed.


 \section{The geometric Hamilton--Jacobi theory}
\label{firstsec}

We summarize the main features of the geometric Hamilton--Jacobi theory
 for the Hamiltonian and Lagrangian formalisms
of autonomous dynamical systems as it is stated in
\cite{art:Carinena_Gracia_Marmo_Martinez_Munoz_Roman06}
(see also  \cite{Ab,KV-1993}).

 \subsection{Hamiltonian Hamilton--Jacobi problem}

Typically, a {\sl (regular autonomous)  Hamiltonian system} is a triad
$(\Tan^*Q,\omega,H )$, where the bundle
$\pi_Q\colon\Tan^*Q\to Q$ represents the {\sl phase space} of a dynamical system
($Q$ is the {\sl configuration space\/}),
$\omega=-\d\theta \in\df^2(\Tan^*Q)$ is the natural symplectic form in $\Tan^*Q$,
and $H \in \Cinfty (\Tan^*Q)$ is the {\sl Hamiltonian function}.
The dynamical trajectories are the integral curves 
$\sigma \colon I\subseteq\Real\to \Tan^*Q$
of the {\sl Hamiltonian vector field} $Z_H\in\vf(\Tan^* Q)$ associated with $H$,
which is the solution to the {\sl Hamiltonian equation}
\beq
\inn(Z_H)\omega=\d H \ .
\label{hameq}
\eeq
(Here, $\df^k(\Tan^*Q)$ and $\vf(\Tan^* Q)$ are the sets of differentiable $k$-forms and vector fields in $\Tan^*Q$, and $\inn(Z_H)\omega$ denotes the inner contraction of $Z_H$ and $\omega$). 
In a chart of natural coordinates $(q^i,p_i)$ in $\Tan^*Q$
we have that $\omega=\d q^i\wedge\d p_i$,
and the curves $\sigma(t)=(q^i(t),p_i(t))$
are the  solution to the {\sl Hamilton equations}
$$
\frac{\d q^i}{\d t} = \derpar{H}{p_i}(q(t),p(t)) \quad , \quad 
\frac{\d p_i}{\d t} = -\derpar{H}{q^i}(q(t),p(t)) \ .
$$

\begin{definition}
\label{genHJh}
The  \textbf{generalized Hamiltonian Hamilton--Jacobi problem}
for a Hamiltonian system $(\Tan^*Q,\omega,H )$
is to find a vector field $X\in\vf(Q)$ and
a $1$-form $\alpha\in\df^1(Q)$
such that, if $\gamma\colon\Real\to Q$ is an integral curve
of $X$, then $\alpha\circ\gamma\colon\Real\to\Tan^*Q$
is an integral curve of $Z_H$; that is,
if $X\circ\gamma=\dot\gamma$, then
$\dot{\overline{\alpha\circ\gamma}}=Z_H\circ(\alpha\circ\gamma)$.
Then, the couple $(X,\alpha)$ is a  \textbf{solution to the generalized Hamiltonian Hamilton--Jacobi problem}.
\end{definition}

\begin{teor}
\label{HJhcns}
The following statements are equivalent:
\begin{enumerate}
\item
The couple $(X,\alpha)$ is a solution to the generalized Hamiltonian Hamilton--Jacobi problem.
\item 
The vector fields $X$ and $Z_H$ are $\alpha$-related; that is,
$Z_H\circ\alpha=\Tan\alpha\circ X$.
As a consequence
$X=\Tan\pi_Q\circ Z_H\circ\alpha$ and it is called the
{\rm vector field associated with the form $\alpha$}.
\item
The submanifold ${\rm Im}\,\alpha$ of $\Tan^*Q$
is invariant by the Hamiltonian vector field $Z_H$
(or, what means the same thing, $Z_H$ is tangent to ${\rm Im}\,\alpha$).
\item
The integral curves of $Z_H$ which have their initial conditions
in $\mathrm{Im}\,\alpha$ project onto the integral curves of 
$X$.
\item
The equation
\ $\inn(X)\d\alpha=-\d(\alpha^*H)$ holds for the $1$-form $\alpha$.
\end{enumerate}
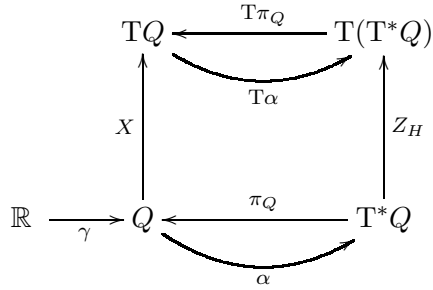
$$
\xymatrix{
\ & \Tan Q \ar@/_1.5pc/[rr]_{\Tan\alpha}  & \ & \ar[ll]_{\Tan\pi_Q} \Tan(\Tan^*Q) \\
\ & \ & \ & \ \\
\Real  \ar[r]_{\gamma} \ & Q \ar[uu]^{X} \ar@/_1.5pc/[rr]_{\alpha} & \ & \ar[ll]_{\pi_Q} \Tan^*Q \ar[uu]_{Z_H}
}
$$
\end{teor}
\proof
(Guidelines for the proof):

The equivalence between 1 and 2 is a consequence of the Definition \ref{genHJh} and the definition of integral curves.
Then, the expression $X=\Tan\pi_Q\circ Z_H\circ\alpha$ is obtained by composing both members of the equality 
$Z_H\circ\alpha=\Tan\alpha\circ X$ with $\Tan\pi_Q$
and taking into account that $\pi_Q\circ\alpha={\rm Id}_Q$.

Items 3 and 4 follow from 2.

Item 5 is obtained from Definition \ref{genHJh} and using the dynamical equation \eqref{hameq}.
\qed

In order to solve the generalized Hamilton--Jacobi problem,
it is usual to state a less general version of it,
which constitutes the standard Hamilton--Jacobi problem.

\begin{definition}
\label{HJh}
The \textbf{Hamiltonian Hamilton--Jacobi problem}
for a Hamiltonian system $(\Tan^*Q,\omega,H )$ 
is to find a $1$-form $\alpha\in\df^1(\Tan^*Q)$
such that it is a solution to the generalized Hamiltonian Hamilton--Jacobi problem and is closed.
Then, the form  $\alpha$ is a \textbf{solution to the Hamiltonian Hamilton--Jacobi problem}.
\end{definition}

As $\alpha$ is closed, for every point in $Q$,
there is a function $S$ in  a neighbourhood $U\subset Q$
such that $\alpha=\d S$.
It is called a {\sl local generating function} of the solution $\alpha$.

\begin{teor}
\label{cnsh}
The following statements are equivalent:
\begin{enumerate}
\item
The form  $\alpha\in\df^1(Q)$ is a solution to the Hamiltonian Hamilton--Jacobi problem.
\item
$\mathrm{Im}\,\alpha$ is a Lagrangian submanifold of
$\Tan^*Q$ which is invariant by $Z_H$,
and $S$ is a local generating function of this Lagrangian submanifold.
\item
The equation $\d(\alpha^*H)=0$ holds for $\alpha$ 
or, what is equivalent, the function
$H\circ\d S\colon Q\to\Real$ is locally constant.
\end{enumerate}
\end{teor}
\proof
(Guidelines for the proof):
They are consequences of Theorem \ref{HJhcns} and Definition \ref{HJh}.
\qed

The last condition, written in natural coordinates, gives the classical form of the
Hamiltonian Hamilton--Jacobi equation, which is
\beq
H\left(q^i,\derpar{S}{q^i}\right)= E\ (ctn.) \ .
\label{tres}
\eeq

These forms $\alpha$ are particular solutions to the (generalized) Hamilton--Jacobi problem, 
but we are also interested in finding {\sl complete solutions} to the problem.
Then:

\begin{definition}
\label{completeHJ}
Let $\Lambda\subseteq\Real^n$.
A family of  solutions $\{ \alpha_\lambda; \lambda\in\Lambda\}$,
 depending on $n$ parameters
$\lambda\equiv (\lambda_1,\ldots,\lambda_n)\in\Lambda$, 
is a \textbf{complete solution} to the Hamiltonian Hamilton--Jacobi problem
if the map 
$$
\begin{array}{ccccc}
\Phi & \colon &Q\times\Lambda & \longrightarrow & \Tan^*Q \\
 & & (q,\lambda) & \mapsto & \alpha_\lambda(q)
\end{array}
$$
is a local diffeomorphism.
\end{definition}

\begin{remark}{\rm
Given a complete solution $\{ \alpha_\lambda; \lambda\in\Lambda\}$,
as $\d\alpha_\lambda=0$, $\forall\lambda\in\Lambda$, 
there is a family of functions $\{ S_\lambda\}$ defined in 
neighbourhoods $U_\lambda\subset Q$ of every point
such that $\alpha_\lambda=\d S_\lambda$.
Therefore we have a locally defined function
$$
\begin{array}{ccccc}
{\cal S} & \colon &\bigcap U_\lambda\times\Lambda\subset Q\times\Lambda & \longrightarrow & \Real \\
 & & (q,\lambda) & \mapsto & S_\lambda(q)
\end{array}
$$
which is called a {\sl local generating function} of the 
complete solution $\{ \alpha_\lambda; \lambda\in\Lambda\}$.

\noindent  A complete solution defines a Lagrangian foliation in $\Tan^*Q$
which is  transverse to the fibers,
and such that $Z_H$ is tangent to the leaves.
The functions that define locally this foliation are the components of a map
$F\colon \Tan^*Q\stackrel{\Phi^{-1}}{\longrightarrow} Q\times\Lambda 
\stackrel{pr_2}{\longrightarrow} \Lambda\subset\Real^n$,
and give a family of constants of motion of $Z_H$.
Conversely, if we have $n$ first integrals $f_1,\ldots,f_n$ 
of $Z_H$ in involution, such that $\d f_1\wedge\ldots\wedge \d f_n\not=0$;
then $f_i=\lambda_i$, with $\lambda_i\in\Real$,
define this transversal Lagrangian foliation and hence a
local complete solution $\{\alpha_\lambda,\lambda\in\Lambda\}$.
Thus we can  locally isolate
$p_i=p_i(q,\lambda)$, replace them in $Z_H$
and project to the basis, then obtaining the family of vector fields 
$\{X_\lambda\}$ associated with the local complete solution.
If $\{ \alpha_\lambda;\lambda\in\Lambda\}$ is a complete solution,
then all the integral curves of  $Z_H$ are obtained from
the integral curves of the associated vector fields $\{ X_\lambda\}$.}
\end{remark}


\subsection{Lagrangian Hamilton--Jacobi problem}

The above framework for the Hamilton--Jacobi theory can
be easily translated to the Lagrangian formalism of mechanics.
Now, the phase space is
the tangent bundle $\tau_Q\colon\Tan Q\to Q$ of the configuration bundle $Q$ and
the dynamics is given by the Lagrangian function of the system,
$\L\in\Cinfty(\Tan Q)$.
Using the canonical structures in $\Tan Q$; that is,
the {\sl vertical endomorphism} $S\in{\mathfrak T}^1_1(\Tan Q)$,
and the {\sl Liouville vector field} $\Delta\in\vf(\Tan Q)$,
the Lagrangian forms
$\theta_\L:=\d\L\circ S\in\df^1(\Tan Q)$, $\omega_\L=-d\theta_\L\in\df^2(\Tan Q)$,
and the Lagrangian energy
$E_\L:=\Delta(\L)-\L\in\Cinfty(\Tan Q)$ are constructed.
Then the {\sl Lagrangian equation} is
\begin{equation}
\inn(\Gamma_\L)\omega_\L=\d E_\L \ ,
\label{ELeq}
\end{equation}
and $(\Tan Q,\omega_\L,E_\L)$ is a {\sl Lagrangian dynamical system}.
Furthermore, the {\sl Legendre transformation} associated with $\L$,
denoted by $\FL\colon \Tan Q\to \Tan^*Q$, is defined
as the fiber  derivative of the Lagrangian function.
We assume that $\L$ is regular; that is,
$\FL$ is a local diffeomorphism or, equivalently,
$\omega_\L$ is a symplectic form
(the Lagrangian is {\sl hyper-regular} if $\FL$ is a global diffeomorphism).
In that case, the Lagrangian equation (\ref{ELeq})
has a unique solution $\Gamma_\L\in\vf(\Tan Q)$,
which is called the {\sl Lagrangian vector field},
whose integral curves are holonomic ,
and are the solutions to the Euler-Lagrange equations.
(See \cite{Cr-83} for details).

\begin{definition}
\label{genHJl}
The  \textbf{generalized Lagrangian Hamilton--Jacobi problem} 
for a Lagrangian system $(\Tan Q,\omega_\L,E_\L)$ 
is to find a vector field $X\in\vf(Q)$
such that, if $\gamma\colon\Real\to Q$ is an integral curve
of $X$, then $X\circ\gamma=\dot\gamma\colon\Real\to \Tan Q$
is an integral curve of $\Gamma_\L$; that is,
if $X\circ\gamma=\dot\gamma$, then
$\Gamma_\L\circ\dot\gamma=\dot{\overline{X\circ\gamma}}$.
Then, the vector field $X$ is a \textbf{solution to the generalized 
Lagrangian Hamilton--Jacobi problem}.
\end{definition}

\begin{teor}
\label{HJlcns}
The following statements are equivalent:
\begin{enumerate}
\item 
The vector field $X$ is a solution to the generalized Lagrangian
Hamilton--Jacobi problem.
\item
The vector fields $X$ and $\Gamma_\L$ are $X$-related; that is,
$\Gamma_\L\circ X=TX\circ X$.
\item
The submanifold ${\rm Im}\,X$ of $\Tan Q$
is invariant by the Lagrangian vector field $\Gamma_\L$
(or, what means the same thing, $\Gamma_\L$ is tangent to ${\rm Im}\,X$).
\item
The integral curves of $\Gamma_\L$ which have their initial conditions
in ${\rm Im}\, X$ project onto the integral curves of $X$.
\item
The equation
\ $\inn(X)(X^*\omega_\L)=\d(X^*E_\L)$ holds for the vector field $X$.
\end{enumerate}
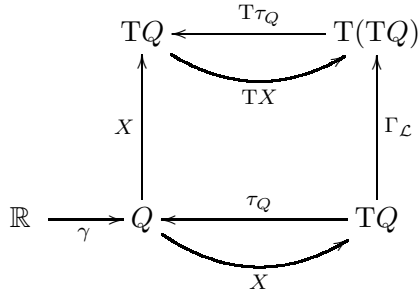
$$
\xymatrix{
\ & \Tan Q \ar@/_1.5pc/[rr]_{\Tan X}  & \ & \ar[ll]_{\Tan\tau_Q} \Tan(\Tan Q) \\
\ & \ & \ & \ \\
\Real  \ar[r]_{\gamma} \ & Q \ar[uu]^{X} \ar@/_1.5pc/[rr]_{X} & \ & \ar[ll]_{\tau_Q} \Tan Q \ar[uu]_{\Gamma_\Lag}
}
$$
\end{teor}
\proof
(Guidelines for the proof):
The proof follows the same patterns as Theorem \ref{HJhcns}.
\qed

As in the Hamiltonian formalism, we consider the following simpler case:

\begin{definition}
\label{HJl}
The  \textbf{Lagrangian Hamilton--Jacobi problem} for a Lagrangian system 
$(\Tan Q,\omega_\L,E_\L)$ is to find a vector field $X$  such that
it is a solution to the generalized Lagrangian Hamilton--Jacobi problem
and satisfies that $X^*\omega_\L=0$.
Then, this vector field $X$ is a \textbf{solution to the Lagrangian Hamilton--Jacobi problem}.
\end{definition}

Since $0=X^*\omega_\L=-X^*\d\theta_\L=-d(X^*\theta_\L)$
then, for every point of $Q$, there is a neighbourhood $U\subset Q$
and a function $S$ such that
$X^*\theta_\L=\d S$ (in $U$).

\begin{teor}
\label{cnsl}
The following statements are equivalent:
\begin{enumerate}
\item 
The vector field $X$ is a solution to the Lagrangian Hamilton--Jacobi problem.
\item
${\rm Im}\,X$ is a Lagrangian submanifold of $\Tan Q$ which
is invariant by the Lagrangian vector field $\Gamma_\L$
(and $S$ is a local generating function of this Lagrangian submanifold).
\item
The equation
$\d(X^*E_\L)=0$ holds for $X$ or, what is equivalent, the function
$E_\Lag\circ\d S\colon Q\to\Real$ is locally constant.
\end{enumerate}
\end{teor}
\proof
(Guidelines for the proof):
They are consequences of Theorem \ref{HJlcns} and Definition \ref{HJl}.
\qed

The last condition leads to the following expression
which is the form of the Lagrangian Hamilton--Jacobi equation in natural coordinates,
\beq
\label{eqn:LagHJCharFunctLocalClassic}
\derpar{S}{q^i} = \derpar{\Lag}{v^i}(q^i,X^i) \ .
\eeq

As in the Hamiltonian Hamilton--Jacobi theory,
we are interested in the complete solutions to the problem,
which are defined as:

\begin{definition}
\label{lagcomplete}
Let $\Lambda\subseteq\Real^n$.
A family of  solutions$\{ X_\lambda; \lambda\in\Lambda\}$
depending on $n$ parameters 
$\lambda\equiv (\lambda_1,\ldots,\lambda_n)\in\Lambda$, 
is a \textbf{complete solution} to the Lagrangian Hamilton--Jacobi problem
if the map 
$$
\begin{array}{ccccc}
\Psi & \colon &Q\times\Lambda & \longrightarrow & \Tan Q \\
 & & (q,\lambda) & \mapsto & X_\lambda(q)
\end{array}
$$
is a local diffeomorphism.
\end{definition}

If we have a complete solution to the Lagrangian Hamilton--Jacobi problem,
all the integral curves of the Lagrangian vector field $\Gamma_\L$
are obtained from the integral curves of all the vector fields$X_\lambda$.

The equivalence between the Lagrangian and the Hamiltonian 
Hamilton--Jacobi problems is stated as follows:

\begin{teor}
\label{relation}
Let \ $(\Tan Q,\omega_\L,E_\L)$ \ be a (hyper)regular Lagrangian system, and
$(\Tan^*Q,\omega,H)$ its associated Hamiltonian system.
If $\alpha\in\df^1(Q)$ is a solution to the 
(generalized) Hamiltonian Hamilton--Jacobi
problem, then $X=\FL^{-1}\circ\alpha$ is a solution to the
(generalized) Lagrangian Hamilton--Jacobi problem
and conversely,
If $X\in\vf(Q)$ is a solution to the
(generalized) Lagrangian Hamilton--Jacobi
problem, then $\alpha=\FL\circ X$ is a solution to the
(generalized) Hamiltonian Hamilton--Jacobi problem.
\end{teor} 
\proof
(Guidelines for the proof):
It can be proven that $\alpha={\cal FL}\circ X$; then,
bearing in mind that $\Tan{\cal FL}\circ\Gamma_\L=Z_H\circ{\cal FL}$,
the proof follows using items 2 and 5 of Theorems \ref{HJhcns} and \ref{HJlcns} (or item 3 of Theorems \ref{cnsh} and \ref{cnsl}).
\qed


 \section{The ``classical'' Hamilton--Jacobi theory}
\label{clasic}

In this section we review the geometric description
of the classical Hamiltonian Hamilton--Jacobi theory (for autonomous systems),
based on using canonical transformations \cite{Ab,Ar,JS,LM-87,MMMcq}.
It is stated in the Hamiltonian formalism .

 \subsection{Canonical transformations and the classical Hamiltonian Hamilton--Jacobi problem}

First we remind the following well-known results \cite{Ab}:

\begin{prop}
Let $(M_1,\omega_1)$, $(M_2,\omega_2)$ be symplectic manifolds
and $\pi_j\colon M_1\times M_2\to M_j$, $j=1,2$.
Then $(M_1\times M_2,\pi_1^*\omega_1-\pi_2^*\omega_2)$
is a symplectic manifold.
\end{prop}

\begin{prop}
Let $\Phi\colon M_1\to M_2$ be a diffeomorphism and
${\jmath}\colon graph\,\Phi\hookrightarrow M_1\times M_2$.\\
$\Phi$ is a symplectomorphism (i.e., $\Phi^*\omega_2=\omega_1$) if, and only if,
$graph\,\Phi$ is a Lagrangian submanifold of $(M_1\times M_2,\pi_1^*\omega_1-\pi_2^*\omega_2)$.
\end{prop}

If $\omega_j=-\d\theta_j$, $j=1,2$;
being $graph\,\Phi$ a Lagrangian submanifold we have
\beq
0=\jmath^*(\pi_1^*\omega_1-\pi_2^*\omega_2)=\d\jmath^*(\pi_2^*\theta_2-\pi_1^*\theta_1)
\ \Longleftrightarrow\ \jmath^*(\pi_2^*\theta_2-\pi_1^*\theta_1)\vert_{\cal W}=-\d{\cal S} \ .
\label{Wlocal}
\eeq
${\cal S}$ is a function defined in 
an open neighbourhood ${\cal W}\subset graph\,\Phi$ of every point, which 
depends on the choice of $\theta_1$ and $\theta_2$.

\begin{definition}
${\cal S}$ is called a
\textbf{generating function} of the Lagrangian submanifold $graph\,\Phi$
and hence of the symplectomorphism $\Phi$.
\end{definition}

If  $({\cal U}_1; q^i,p_i)$, $({\cal U}_2; \tilde q^i,\tilde p_i)$ are Darboux charts such that
${\cal W}\subset {\cal U}_1\times {\cal U}_2$, local coordinates in ${\cal W}$  can be chosen in several ways.
This leads to different possible choices for ${\cal S}$.
Thus, for instance,  if $({\cal W}; q^i\tilde q^i)$ is a chart,
then \eqref{Wlocal} gives the symplectomorphism  explicitly as
$$
\tilde p_i\d\tilde q^i-p_i\d q^i=-\d {\cal S}(q,\tilde q)
\quad \Longleftrightarrow\ \quad
\tilde p_i=-\derpar{{\cal S}}{\tilde q^i}(q,\tilde q)\ , \ p_i=\derpar{{\cal S}}{q^i}(q,\tilde q)\ .
$$

Now, let $(\Tan^*Q,\omega,H )$ be a Hamiltonian system.

\begin{definition}
A \textbf{canonical transformation} for a Hamiltonian system
$(\Tan^*Q,\omega,H )$ is a symplectomorphism
$\Phi\colon \Tan^*Q\to \Tan^*Q$.
As a consequence, $\Phi$ transforms Hamiltonian vector fields into Hamiltonian vector fields.
\end{definition}


\begin{definition}
The \textbf{Hamilton--Jacobi problem}
for a Hamiltonian system $(\Tan^*Q,\omega,H )$  consists in finding
a canonical transformation $\Phi\colon \Tan^*Q\to \Tan^*Q$ leading the
system to equilibrium; that is, such that
$H\circ\Phi=E\ (ctn.)$.
\end{definition}
The canonical transformation $\Psi$ is given by a generating function ${\cal S}$:
\beq
\derpar{{\cal S}}{q^i}(q,\tilde q)=p_i \quad , \quad
-\derpar{{\cal S}}{\tilde q^i}(q,\tilde q)=\tilde p_i \ ,
\label{uno}
\eeq
where ${\cal S}$ the general solution to the {\sl Hamilton--Jacobi equation\/}
\beq
H\left(q^i,\derpar{{\cal S}}{q^i}\right)= E\ (ctn.) \ .
\label{unobis}
\eeq
Then, the Hamilton equations for the transformed Hamiltonian function $H\circ\Phi\equiv\tilde H$ are
\beq
\frac{\d\tilde q^i}{\d t} = \derpar{\tilde H}{\tilde p_i}(\tilde q(t),\tilde p(t))=0 \quad , \quad 
\frac{\d\tilde p_i}{\d t} = -\derpar{\tilde H}{\tilde q^i}(\tilde q(t),\tilde p(t))=0 \ ;
\label{dos}
\eeq
and solving \eqref{unobis}, from (\ref{dos}) and (\ref{uno}), the dynamical curves $(q^i(t),p_i(t))$
of the original Hamiltonian system $(\Tan^*Q,\omega,H )$ are obtained.

 \subsection{Relation between the ``classical'' and the geometric Hamilton--Jacobi theories}

The  relation between the ``classical'' and the geometric Hamilton--Jacobi theories
is established through the equivalence of complete solutions 
and canonical transformations (see also \cite{Vi-14}).

\begin{teor}
Let $(\Tan^*Q,\omega,H )$ be a Hamiltonian system.
A complete solution $\{ \alpha_\lambda; \lambda\in\Lambda\}$
to the Hamilton--Jacobi problem provides a canonical transformation 
$\Phi\colon \Tan^*Q\to \Tan^*Q$ leading the system to equilibrium,
and conversely.
\end{teor}
\proof
In  a neighbourhood of every point, consider a complete solution
$\{ \alpha_\lambda; \lambda\in\Lambda\}$, and
let ${\cal S}$ be a generating function of it.
As ${\cal S}={\cal S}(q^i,\lambda^i)$, we can identify $\lambda^i$ 
with a subset  of coordinates $\lambda^i\equiv\tilde q^i$ in $\Tan^*Q\times\Tan^*Q$, and then
${\cal S}={\cal S}(q^i,\tilde q^i)$ can be thought as a generating function
of a  local canonical transformation $\Phi$ and hence
of an open set ${\cal W}$ of the Lagrangian submanifold 
$graph\,\Phi\hookrightarrow \Tan^*Q\times\Tan^*Q$.
Making this construction in every chart, we have the transformation $\Phi$
and the submanifold $graph\,\Psi$.
Now, as (\ref{tres}) holds for every particular solution $S_\lambda$, we have that
$$
E=H\left(q^i(\tilde q,\tilde p),\derpar{{\cal S}}{q^i}(q(\tilde q,\tilde p),\tilde q)\right)= 
\tilde H(\tilde q^i,\tilde p_i) \ .
$$
Conversely, if we have the canonical transformation $\Psi$, from a
generating function ${\cal S}={\cal S}(q^i,\tilde q^i)$, taking
$\tilde q\equiv (\tilde q^i)=(\lambda^i)\equiv\lambda$, we obtain
a family of functions $\{ S_\lambda\}$ and, hence a
local complete solution $\{\alpha_\lambda=\d S_\lambda; \lambda\in\Lambda\}$
to the Hamiltonian Hamilton--Jacobi problem.
Making this construction in every chart, we have the complete solution.

Geometrically, this means that, on each local chart of $\Tan^*Q$,
fixing the coordinates $\tilde q^i=\lambda^i$ of a point,
we obtain a local submanifold whose image by $\Phi^{-1}$ gives
the image of a local section $\alpha_\lambda\colon Q\to \Tan^*Q$
which is a particular solution to the Hamiltonian Hamilton--Jacobi problem.
\qed


\section{General geometric framework for the Hamilton--Jacobi theory}
\protect\label{co}

The geometric Hamilton--Jacobi theory can be stated in a more general framework
which allows us to extend the theory to a wide variety of systems and situations.
Next we present a summary of this general framework as it is stated in \cite{HJTeam3} 
(see also \cite{BMMP-10} for another similar approach).

\subsection{Slicing problems}

In general, a {\sl dynamical system} is just a couple $(P,Z)$,
where $P$ is a manifold and $Z\in\vf(P)$ is a vector field
which defines the dynamical equation on~$P$.
Then, in order to state the analogous to the Hamilton--Jacobi problem for this system
in a more general context, consider a manifold~${\cal M}$,
a vector field $X\in\vf({\cal M})$, and a map $\alpha \colon {\cal M} \to P$,
as it is showed in the following diagram:
$$
\xymatrix{
*++{\Tan {\cal M}}  \ar[r]^{\Tan \alpha}  \ar[d]_{} &
*++{\Tan P}  \ar[d]_{}
\\
*++{{\cal M}} \ar[r]^{\alpha} \ar@/^3mm/[u]^{X} &
*++{P} \ar@/_3mm/[u]_{Z}
}
$$

\begin{prop}
\label{firstprop}
The following statements are equivalent:
\begin{enumerate}
\itemsep 0pt plus 1pt
\item
If $\gamma$ is an integral curve of $X$, then
$\zeta= \alpha\circ\gamma$ is an integral curve of~$Z$.
\item
The vector fields $X$ and $Z$ are $\alpha$-related:
\beq
\label{slicing}
\Tan \alpha\circ X = Z\circ\alpha \ ,
\eeq
\end{enumerate}
Furthermore, if $\alpha$ is an injective immersion,
(inducing a diffeomorphism
$\alpha_o \colon {\cal M} \to \alpha({\cal M})$),
then these properties are equivalent to:
\begin{enumerate}
\itemsep 0pt plus 1pt
\setcounter{enumi}{2}
\item
The vector field $Z$ is tangent to $\alpha({\cal M})$, 
and, if $Z_o= Z|_{\alpha({\cal M})}$,
then $X = \alpha_o^*(Z_o)$.

Then, the map $\gamma\mapsto \alpha\circ\xi$
is a bijection between the integral curves of~$X$
and the integral curves of~$Z$ in $\alpha({\cal M})$.
\end{enumerate}
\end{prop}
\proof
They are immediate, bearing in mind the commutativity of the above diagram.
\qed

\begin{definition}
A \textbf{slicing} of a dynamical system $(P,Z)$ is
a triple $({\cal M},\alpha,X)$ which is a solution to the
\textbf{slicing equation} \eqref{slicing}.
\end{definition}

If $(x^i)$ and $(z^j)$ are coordinates in~${\cal M}$ and $P$, respectively;
and $\alpha(x) = (a^j(x))$, $\ds X = X^i \derpar{}{x^i}$,
and $\ds Z = Z^j\derpar{}{z^j}$, then
$\ds (\Tan \alpha \circ X - Z \circ \alpha)(x^i)=\Big(a^j(x) ,\derpar{a^j}{x^i} \, X^i -Z^j(\alpha(x))\Big)$,
and $({\cal M},\alpha,X)$ is a solution to the slicing equation if, and only if,
$$
\ds \derpar{a^j}{x^i} \, X^i(x) = Z^j (\alpha(x)) \ .
$$

We say that the vector field $X$ gives a ``partial dynamics'' or a ``slice'' 
of the ``whole dynamics'' which is given by $Z$, 
and the whole dynamics can be recovered from these slices.
In fact, the integral curves of $Z$ contained in $\alpha({\cal M}) \subset P$
can be described by a solution $(\alpha,X)$ to the slicing equation; 
but we need a {\sl complete solution} to describe all the integral curves of $Z$ 
and it can be defined as a family of solutions depending on the parameters
of a space $\Lambda\subseteq{\mathbb R}^n$.

\begin{definition}
\label{compslice}
A \textbf{complete slicing} of a dynamical system $(P,Z)$ is a map 
$\overline\alpha\colon {\cal M}\times\Lambda\to P$
and a vector field $\overline X \colon {\cal M} \times\Lambda \to \Tan {\cal M}$
along the projection ${\cal M} \times\Lambda \to {\cal M}$
such that:
\ben
\item
The map $\overline\alpha$ is surjective,
\item
for every $\lambda\in\Lambda$, the map $\alpha_\lambda\colon {\cal M} \to P$
and $X_\lambda\colon {\cal M} \to \Tan {\cal M}$ are a slicing of $Z$.
\een
$$
\xymatrix{
*++{\Tan {\cal M} \times\Lambda} \ar[r]^{\Tan_1 \overline\alpha} \ar[d]_{} &
*++{\Tan P} \ar[d]_{}
\\
*++{{\cal M} \times\Lambda} \ar[r]^{\overline\alpha} \ar@/^3mm/[u]^{\overline X} &
*++{P} \ar@/_3mm/[u]_{Z}
}
$$
\end{definition}

Thus, a complete slicing is a family of maps
$\alpha_\lambda \equiv \overline\alpha(\cdot,\lambda) \colon {\cal M} \to P$
and vector fields
$X_\lambda \equiv \overline X(\cdot,\lambda) \colon {\cal M} \to \Tan {\cal M}$
satisfying the above conditions.

As for every point $p\in P$ there exits $(x,\lambda) \in {\cal M} \times\Lambda$ 
such that $\overline\alpha(x,\lambda)=p$;
the integral curve of $Z$ through~$p$
is described by the integral curve of $X_\lambda$ through $x$
by means of the map $\alpha_\lambda$.
In addition, if each $\alpha_\lambda$ is an immersion
(for instance,when it is a diffeomorphism)
then $X_\lambda$ are determined by the $\alpha_\lambda$.

The hypothesis of $\alpha$ being an embedding holds in many situations;
for instance, for the sections of a fiber bundle $\pi \colon P \to {\cal M}$.
Then we can consider the slicing problem
for sections $\alpha \colon {\cal M} \to P$ of~$\pi$,  as before.
In this case, as $\alpha$ is an embedding,
the equation \eqref{slicing} determines $X$, and
$X$ is given from~$\alpha$ by the equation
$$
X = \Tan \pi \circ Z \circ \alpha \ .
$$
In this case, Proposition \ref{firstprop} states that
a section $\alpha$ of $\pi \colon P \to {\cal M}$
is a solution to the slicing equation for $(P,Z)$ if, and only if,
$$
\Tan \alpha \circ \Tan \pi \circ Z \circ \alpha = Z \circ \alpha \ .
$$

\subsection{Recovering the Hamilton--Jacobi equation for Hamiltonian and Lagrangian dynamical systems}

Consider the case of a Hamiltonian system $(P,\omega,H)$,
where $(P,\omega)$ is a symplectic manifold, $H\in\Cinfty(P)$ is a
Hamiltonian function, and $Z = Z_H$ is its Hamiltonian vector field; that is,
the solution to \eqref{hameq}. Then:

\begin{teor}
\label{prop-ham}
If $({\cal M},\alpha,X)$ is a solution to the slicing equation \eqref{slicing} for $(P,Z_H)$, then
$$
\inn(X)\,\alpha^*\omega - \d \,\alpha^*H = 0 \ .
$$
In addition, if $\alpha \colon {\cal M} \to P$ is an embedding
satisfying the condition $\alpha^*\omega= 0$, then
$$
\d\,(\alpha^*H)= 0 \ ;
$$
and conversely, if $\dim P = 2 \dim {\cal M}$
and $\alpha$ satisfies this equation and $\alpha^*\omega= 0$,
then $\alpha$ is a solution to the slicing equation \eqref{slicing}.
\end{teor}

In the particular case where $\pi \colon P \to {\cal M}$ is a fiber bundle
(for instance, ${\cal M}=Q$ and $P=\Tan^*Q$),
we can consider the slicing problem as before,
but only for sections of~$\pi$,
$$
\xymatrix{
*++{\Tan {\cal M}}  \ar[r]^{\Tan \alpha}  \ar[d]_{} &
*++{\Tan P}  \ar[d]_{}
\\
*++{{\cal M}} \ar[r]^{\alpha} \ar@/^3mm/[u]^{X} &
*++{P} \ar@/^3mm/[l]^{\pi} \ar@/_3mm/[u]_{Z}
}
$$
Being $\alpha$ an embedding, the equation (\ref{slicing}) determines~$X$,
and composing this equation with $\Tan\pi$, we obtain that
$X = \Tan\pi \circ Z_H \circ \alpha$.
Therefore the slicing equation \eqref{slicing} reads
$$
\Tan\alpha \circ\Tan\pi \circ Z \circ\alpha = Z \circ\alpha \ .
$$

In this way, the equation \eqref{slicing} can be considered as a
generalization of the Hamilton--Jacobi equation  in the
Hamiltonian formalism, which
is just the slicing equation for a closed $1$-form~$\alpha$ in $Q$.
Therefore,  $\alpha = \d S$ locally,
and the slicing equation looks in the ordinary form
$H \circ\d S = \mathrm{const}$.

The same applies to the Lagrangian formalism.
In this case $P = \Tan Q$ and, if $\Lag\in\Cinfty(\Tan Q)$ 
is a regular Lagrangian function, $Z=\Gamma_\Lag$
is the Lagrangian vector field solution to the Lagrangian equation \eqref{ELeq}.
Then all proceeds as in the Hamiltonian case.

The definitions \ref{completeHJ} and \ref{lagcomplete} of complete solutions to the 
Hamiltonian and Lagrangian Hamilton--Jacobi problems respectively
are particular cases of the definition \ref{compslice} of complete slicings.


\section{The Hamilton--Jacobi problem for other physical systems}
\label{others}

Using the general framework presented in the above section,
the Hamilton--Jacobi problem can be stated for a wide kind of physical systems.
Next we review two of them.
(Other applications of the theory are listed in detail in the Introduction). 


\subsection{Higher-order (autonomous) dynamical systems}

Let $Q$ be a $n$-dimensional manifold) and let
$\Tan^kQ$ the $k$th-order tangent bundle of $Q$,
which is endowed with natural coordinates
$\left(q_{0}^A,q_{1}^A,\ldots,q_{k}^A\right) = \left(q_i^A\right)$,
$0 \leqslant i \leqslant k$, $1\leqslant A\leqslant n$.
If $\Lag \in\Cinfty(\Tan^{k}Q)$ is the Lagrangian function of
an autonomous $k$th-order Lagrangian system, 
using the canonical structures of the higher-order tangent bundles, 
we can construct the Poincar\'{e}-Cartan
forms and the Lagrangian energy whose coordinate expressions are
\begin{align*}
&\omega_\Lag =-\d \theta_\L=
\sum_{r=1}^k \sum_{i=0}^{k-r}(-1)^{i+1} d_T^i\,\d\left(\derpar{\Lag}{q_{r+i}^A}\right) \wedge \d q_{r-1}^A\in\df^{2}(\Tan^{2k-1}Q) \, , \\
&E_\Lag = \sum_{r=1}^{k}
q_{r}^A \sum_{i=0}^{k-r} (-1)^i d_T^i\left( \derpar{\Lag}{q_{r+i}^A}
\right)- \Lag\in\Cinfty(\Tan^{2k-1}Q) \ ,
\end{align*}
where
$\ds d_Tf\left(q_0^A,\ldots,q_{k+1}^A\right) =
  \sum_{i=0}^{k}q_{i+1}^A \derpar{f}{q_i^A}(q_0^A,\ldots,q_{k}^A)$.
Thus we have the higher-order Lagrangian system $(\Tan^{2k-1}Q,\omega_\Lag,E_\Lag)$.
Assuming that the Lagrangian function is regular; that is,
$\omega_\Lag$ is a symplectic form;
the Lagrangian equation
$\inn(X_\Lag) \, \omega_\Lag = \d E_\Lag$
has a unique solution $X_\Lag \in \vf(\Tan^{2k-1}Q)$ (the Lagrangian vector field)
whose integral curves are holonomic (that is, they are canonical liftings 
$j^{2k-1}\phi\colon\Real\to\Tan^{2k-1}Q$
of curves $\phi\colon\Real\to Q$)
and are the solutions to the {\sl Otrogradskii} or 
{\sl higher-order Euler-Lagrange equations}
(see
\cite{book:DeLeon_Rodrigues85,art:Gracia_Pons_Roman91,art:Prieto_Roman11}
for details).

The Hamilton--Jacobi problem for higher-order 
Lagrangian dynamical systems is just the {\sl slicing problem}
for the particular situation represented in the diagram
$$
\xymatrix{
\Tan(\Tan^{k-1}Q) \ar@/_1.5pc/[rr]_{\Tan s}  & \ & \ar[ll]_{\Tan\rho^{2k-1}_{k-1}} \Tan(\Tan^{2k-1}Q) \\
\ & \ & \ \\
\Tan^{k-1}Q \ar[uu]^{X} \ar@/_1.5pc/[rr]_{s} & \ & \ar[ll]_{\rho^{2k-1}_{k-1}} \Tan^{2k-1}Q \ar[uu]_{X_\Lag}
}
$$
that is, for sections of the natural projection
$\rho^{2k-1}_{k-1}\colon\Tan^{2k-1}Q\to\Tan^{k-1}Q$,
$s \in \Gamma(\rho^{2k-1}_{k-1})$;
and thus we have the following settings
(see \cite{CLRP-13,CLRP-14} for the details and proofs):

\begin{definition}
\label{def:GenLagHJDef}
The \textbf{generalized $k$th-order Lagrangian Hamilton--Jacobi problem}
for the higher-order Lagrangian system $(\Tan^{2k-1}Q,\omega_\Lag,E_\Lag)$
is to find a section $s \in \Gamma(\rho^{2k-1}_{k-1})$ 
and a vector field $X \in \vf(\Tan^{k-1}Q)$ such that,
if $\gamma ~\colon~\R~\to~\Tan^{k-1}Q$ is an integral curve of $X$,
then $s\circ\gamma \colon \R \to \Tan^{2k-1}Q$ is an integral curve of $X_\Lag$;
that is, if $X \circ \gamma = \dot \gamma$, then $X_\Lag \circ (s \circ \gamma) = \dot{\overline{s \circ \gamma}}$.
Then, the couple $(s,X)$ is a \textbf{solution to the generalized  $k$th-order
Lagrangian Hamilton--Jacobi problem}.
\end{definition}

\begin{teor}
\label{higher-gen-cns}
The following statements are equivalent:
\begin{enumerate}
\item 
The couple $(s,X)$ is a solution to the generalized $k$th-order Lagrangian
Hamilton--Jacobi problem.
\item
The vector fields $X$ and $X_\Lag$ are $s$-related; that is,
$X_\Lag \circ s = \Tan s \circ X$.
As a consequence, $X = \Tan\rho^{2k-1}_{k-1} \circ X_\Lag \circ s$,
and $X$ is said to be the {\rm vector field associated with the section $s$}.
\item 
The submanifold ${\rm Im}(s) $ of $\Tan^{2k-1}Q$
is invariant by the Lagrangian vector field $X_\Lag$ 
(or, what means the same thing, $X_\Lag$ is tangent to $s(\Tan^{k-1}Q)$).
\item 
The integral curves of $X_\Lag$ which have initial conditions in $\textnormal{Im}(s)$ 
project onto the integral curves of $X$.
\item 
The equation
$\inn(X)(s^*\omega_\Lag) = \d(s^*E_\Lag)$ holds for $\alpha$.
\end{enumerate}
\end{teor}
\proof
(Guidelines for the proof):
The proof follows a pattern similar to that of Theorem \ref{HJhcns},
but now using Definition \ref{def:GenLagHJDef}.
\qed

\begin{definition}
\label{higherHJl}
The \textbf{$k$th-order Lagrangian Hamilton--Jacobi problem} 
for the higher-order Lagrangian system
$(\Tan^{2k-1}Q,\omega_\Lag,E_\Lag)$ is to find
a section $s \in \Gamma(\rho^{2k-1}_{k-1})$ such that it is a solution to the generalized
$k$th-order Lagrangian Hamilton--Jacobi problem and satisfies that
$s^*\omega_\Lag = 0$. 
Then, this section $s$ is a \textbf{solution to the $k$th-order Lagrangian Hamilton--Jacobi problem}.
\end{definition}

Observe that that $0=s^*\omega_\Lag = -s^*(\d\theta_\Lag) =- \d(s^*\theta_\Lag) = 0$;
that is, $s^*\theta_\Lag$ is a closed $1$-form and then
there exists $S \in \Cinfty(U)$, $U \subset \Tan^{k-1}Q$, such that $s^*\theta_\Lag\vert_U = \d S$.

\begin{teor}
The following statements are equivalent:
\begin{enumerate}
\item 
The section $s$ is a solution to the generalized $k$th-order Lagrangian
Hamilton--Jacobi problem.
\item 
${\rm Im}(s)$ is a Lagrangian submanifold of $\Tan^{2k-1}Q$,
which is invariant by the Lagrangian vector field $X_\Lag$
(and $S$ is a local generating function of this Lagrangian submanifold).
\item 
The equation $\d(s^*E_\Lag) = 0$ holds for $s$
or, what is equivalent, the function
$E_\Lag\circ\d S\colon Q\to\Real$ is locally constant.
\end{enumerate}
\end{teor}
\proof
(Guidelines for the proof):
They are consequences of Theorem \ref{higher-gen-cns} and Definition \ref{higherHJl}.
\qed

In natural coordinates, from this last condition we obtain that
$$
\derpar{S}{q_i^A} =
\sum_{l=0}^{k-i-1}(-1)^ld_T^l\left( \derpar{\Lag}{q_{i+1+l}^A} \right)\Big\vert_{{\rm Im}(s)} \ .
$$
This system of $kn$ partial differential equations for $S$
generalizes the equation \eqref{eqn:LagHJCharFunctLocalClassic}
to higher-order systems.

\begin{definition}
Let $\Lambda\subseteq\Real^n$.
A family of  solutions$\{ s_\lambda; \lambda\in\Lambda\}$,
depending on $n$ parameters 
$\lambda\equiv (\lambda_1,\ldots,\lambda_n)\in\Lambda$,  is a  
\textbf{complete solution to the $k$th-order Lagrangian Hamilton--Jacobi problem} 
if the map 
$$
\begin{array}{ccccc}
\Phi & \colon &\Tan^{k-1}Q\times\Lambda & \longrightarrow & \Tan^{2k-1}Q \\
 & & (q,\lambda) & \mapsto & s_\lambda(q)
\end{array}
$$
is a local diffeomorphism.
\end{definition}

For the Hamiltonian formalism, let 
$h \in \Cinfty(\Tan^*(\Tan^{k-1}Q))$ be the Hamiltonian function 
of a (regular) higher-order dynamical system. 
Using the canonical Liouville forms of the cotangent bundle,
$\theta_{k-1} = p_{A}^{i}\d q_{i}^{A} \in
\df^{1}(\Tan^*(\Tan^{k-1}Q))$ and $\omega_{k-1} = \d q_{i}^{A}\wedge
\d p_{A}^{i} \in \df^{2}(\Tan^*(\Tan^{k-1}Q))$, where
$(q_{i}^{A},p_{i}^{A})$ ($1\leq A\leq n,$ $0\leq i\leq k-1$) are
canonical coordinates in $\Tan^{*}(\Tan^{k-1}Q)$; the
dynamical equation for the Hamiltonian system 
$(\Tan^{*}(\Tan^{k-1}Q),\omega_k,h))$ is
$\inn(X_h)\,\omega_{k-1} = \d h$,
and it has a unique solution $X_h \in \vf(\Tan^*(\Tan^{k-1}Q))$.
As we are working in the cotangent bundle $\Tan^*(\Tan^{k-1}Q)$,
 the Hamiltonian Hamilton--Jacobi problems for higher-order
systems is stated in the same way as in the first-order case
and, hence, it is  the {\sl slicing problem} for the particular situation
represented in the diagram
$$
\xymatrix{
\Tan(\Tan^{k-1}Q) \ar@/_1.5pc/[rrr]_{\Tan \alpha} & \ & \ &
 \ar[lll]_-{\Tan\pi_{\Tan^{k-1}Q}} \Tan(\Tan^*(\Tan^{k-1}Q)) \\
\ & \ & \ & \ \\
\Tan^{k-1}Q \ar[uu]^{X} \ar@/_1.5pc/[rrr]_{\alpha} & \ & \ &
 \ar[lll]_{\pi_{\Tan^{k-1}Q}} \Tan^*(\Tan^{k-1}Q) \ar[uu]_{X_h}
}
$$
Therefore, all the definitions and results are like in the first-order case,
and the relation between both the Lagrangian and the Hamiltonian Hamilton--Jacobi problems
is stated as in Theorem \ref{relation}.

\subsection{Multisymplectic field theories}

The Hamilton--Jacobi theory for multisymplectic field theories
has been studied in \cite{LMM-09,LPRV-2017,art:Vitagliano12}.
Next we state the Lagrangian and the Hamiltonian problems for these systems.
(For details on multisymplectic field theories see, for instance, \cite{proc:Carinena_Crampin_Ibort91,EMR-96,art:Roman09} and the references therein).

\subsubsection{Multisymplectic Lagrangian Hamilton--Jacobi problem}

Let $\pi\colon E\to M$ a bundle, 
where $M$ is an oriented manifold with $\dim\,M=m$ and $\dim\,E=n+m$.
The Lagrangian description of multisymplectic classical field theories
is stated in the first-order jet bundle $\pi^1\colon J^1\pi\to E$,
which is also a bundle $\bar\pi^1\colon J^1\pi\longrightarrow M$.
Natural coordinates in $J^1\pi$ adapted to the bundle structure are 
$(x^i,y^\alpha,y^\alpha_i)$ ($i= 1,\ldots,m$; $\alpha= 1,\ldots,n$).
Giving a Lagrangian density associated to a Lagrangian function $\Lag$
and using the canonical structures of $J^1\pi$
we can define the {\sl Poincar\'e--Cartan forms}
associated with~$\Lag$,
$\Theta_{\Lag}\in\df^{m}(J^1\pi)$ and
$\Omega_{\Lag}:= -\d\Theta_{\Lag}\in\df^{m+1}(J^1\pi)$,
whose local expression is
$$
\Omega_{\Lag}=-\d\Theta_\L=
-\d\left(\derpar{L}{y_i^\alpha}\d y^\alpha \wedge \d^{m-1}x_i -
\Big( \derpar{L}{y_i^\alpha}y_i^\alpha - L \Big)\d^{m}x \right)  \ ,
$$
where $\d^mx = \d x^1 \wedge \ldots \wedge \d x^m$ and 
$\ds\d^{m-1}x_i = \inn\Big(\derpar{}{x^i}\Big)\d^m x$.
The Lagrangian function is regular if $\Omega_{\Lag}$ is a multisymplectic
$(m+1)$-form (i.e., $1$-non\-degenerate).
Then the couple $(J^1\pi,\Omega_\Lag)$ is a {\sl multisymplectic Lagrangian system}.
The {\sl Lagrangian problem} consists in finding $m$-dimensional, 
$\bar{\pi}^1$-transverse, and holonomic distributions 
${\cal D}_\L$ in $J^1\pi$ such that their integral sections
$\psi_\L \in \Gamma(\bar{\pi}^1)$ are canonical liftings $j^1\phi$ of sections 
$\phi\in\Gamma(\pi)$ that are solutions to the
Lagrangian field equation
\beq
(j^1\phi)^*\inn (X)\Omega_\Lag= 0\, , \ \mbox{\textit{for every} } X \in \vf(J^1\pi)  \ .
\label{fieldlag}
\eeq
In coordinates, the components of
$j^1\phi {=}\Big(x^i ,y^\alpha,\ds\derpar{y^\alpha}{x^i}\Big)$
satisfy the {\sl Euler--Lagrange equations}
$$
 \derpar{\Lag}{y^A}\circ j^1\phi-
\derpar{}{x^\mu}\Big(\derpar{\Lag}{y_\mu^A}\circ j^1\phi\Big)= 0   .
$$

\begin{definition}
\label{multigenHJl}
The \textbf{generalized Lagrangian Hamilton--Jacobi problem} 
for the multisymplectic Lagrangian system $(J^1\pi,\Omega_\L)$ 
is to find a section
$\Psi \in \Gamma(\pi^1)$ (which is called a {\sl jet field\/}) 
and an $m$-dimensional integrable distribution ${\cal D}$ in $E$ such that, if
$\gamma \in \Gamma(\pi)$ is an integral section of ${\cal D}$, then
$\psi_\L=\Psi \circ \gamma \in \Gamma(\bar{\pi}^1)$ 
is an integral section of ${\cal D}_\Lag$;
that is, if
$\Tan_{u}{\rm Im}(\gamma) = {\cal D}_{u}$, for every $u \in {\rm Im}(\gamma)$,
then
$\Tan_{\bar{u}}{\rm Im}(\Psi \circ \gamma) = ({\cal D}_\Lag)_{\bar{u}}$,
for every $\bar{u} \in {\rm Im}(\Psi \circ \gamma)$.
Then, the couple $(\Psi,{\cal D})$ is a \textbf{solution to the generalized 
Hamiltonian Hamilton--Jacobi problem}.
\end{definition}

\begin{remark}{\rm
The Hamilton--Jacobi problem can also be stated
associating the distributions ${\cal D}$ and  ${\cal D}_\L$ with {\sl multivector fields}.
An \textsl{$m$-multivector field}, on a manifold ${\cal M}$ is a section of the bundle 
$\Lambda^m(\Tan {\cal M}) \to {\cal M}$,
where $\Lambda^m(\Tan {\cal M})=\Tan {\cal M}\wedge^{\,(m)}_{\ldots\ldots}\wedge\Tan {\cal M}$
 (i.e, a skew-symmetric contravariant tensor field). 
If ${\bf X}$ is an $m$-multivector field in ${\cal M}$ then,
for every $p\in{\cal M}$, there is a neighbourhood 
$U_p \subset {\cal M}$ and local vector fields
$X_1,\ldots,X_m \in \vf(U_p)$ such that 
${\bf X}\vert_{U_p}=X_1\wedge^{\,(m)}_{\ldots\ldots}\wedge X_m$.
Then, if ${\cal D}$ is an $m$-dimensional distribution in ${\cal M}$, 
sections of $\Lambda^m{\cal D} \to {\cal M}$ 
are $m$-multivector fields in ${\cal M}$, and a multivector field is
{\sl integrable} if its associated distribution is also.

\noindent 
Now, if ${\cal M}=J^1\pi$, let ${\bf X}$ and ${\bf X}_\L$ be the $m$-multivector fields
associated with the distributions ${\cal D}$ and  ${\cal D}_\L$, respectively;
then the Lagrangian Hamilton--Jacobi problem can be represented by the diagram
$$
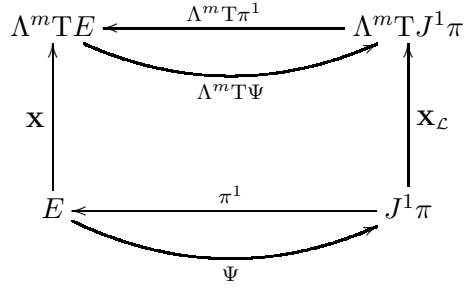
\xymatrix{
\Lambda^m\Tan E \ar@/_1.5pc/[rrr]_{\Lambda^m\Tan\Psi} & \ & \ &
 \ar[lll]_-{\Lambda^m\Tan\pi^1} \Lambda^m\Tan J^1\pi \\
\ & \ & \ & \ \\
E \ar[uu]^{{\bf X}} \ar@/_1.5pc/[rrr]_{\Psi} & \ & \ &
 \ar[lll]_{\pi^1}J^1\pi \ar[uu]_{{\bf X}_\L}
}
$$
(where $\Lambda^m\Tan\Psi$ and $\Lambda^m\Tan\pi^1$ denote the natural extensions of the maps $\psi$ and $\pi^1$ to the multitangent bundles),
and thus this problem can be considered as a special case of a {\sl slicing problem}.
}
\end{remark}

\begin{teor}
\label{prop:LagGenHJEquivalences}
The following statements are equivalent:
\begin{enumerate}
\item 
The couple $(\Psi,{\cal D})$ is a solution to the generalized Lagrangian Hamilton--Jacobi problem.
\item
The distributions ${\cal D}$ and ${\cal D}_\Lag$ are $\Psi$-related.
As a consequence,
${\cal D} = \Tan\pi^1({\cal D}_\L\vert_{{\rm Im}(\Psi)})$, and is called the
{\rm distribution associated with $\Psi$}.
\item 
The distribution ${\cal D}_\Lag$ is tangent to the submanifold ${\rm Im}(\Psi)$
of $J^1\pi$.
\item
The integral sections of ${\cal D}_\Lag$ which have boundary conditions in
${\rm Im}(\Psi)$ project onto the integral sections of  ${\cal D} $.
\item 
If $\gamma$ is an integral section of the distribution
${\cal D}$ associated with the jet field $\Psi$ then, for every $Y \in \vf(E)$, the equation $\gamma^*\inn(Y)(\Psi^*\Omega_\Lag) = 0$
holds for $\Psi$.
\end{enumerate}
\end{teor}
\proof
(Guidelines for the proof):

The equivalence between 1 and 2 is a consequence of the Definition \ref{multigenHJl}, the equivalence between distributions and multivector fields, and the definition of integral sections.

Items 3 and 4 follow from 2.

Item 5 is obtained from Definition \ref{multigenHJl} and using the field equation \eqref{fieldlag}.
\qed

\begin{definition}\label{def:LagHJDef}
The \textbf{Lagrangian Hamilton--Jacobi problem} 
for the multisymplectic Lagrangian system $(J^1\pi,\Omega_\L)$ 
is to find a jet field
$\Psi \in \Gamma(\pi^1)$ such that it is solution to the generalized Lagrangian Hamilton--Jacobi problem and
satisfies that $\Psi^*\Omega_\Lag = 0$. 
Then, the jet field $\Psi$ is a \textbf{solution to the Lagrangian Hamilton--Jacobi problem}.
\end{definition}

The condition $\Psi^*\Omega_\Lag= -\d(\Psi^*\Theta_\Lag)=0$ 
is equivalent to ask that the form 
$\Psi^*\Theta_\Lag $ is closed and 
then there exists a $(m-1)$-form $\omega \in \df^{m-1}(U)$, with $U \subset E$, 
such that $\Psi^*\Theta_\Lag=\d\omega$. Furthermore, 
$\omega$ is $\pi$-semibasic, since $\Theta_\Lag$, 
and hence $\Psi^*\Theta_\Lag$, are also.

\begin{teor}
The following statements are equivalent:
\begin{enumerate}
\item 
The jet field $\Psi$ is a solution to the Lagrangian Hamilton--Jacobi problem.
\item 
${\rm Im}(\Psi)$ is an $m$-Lagrangian submanifold of $J^1\pi$
and the distribution ${\cal D}_\Lag$ is tangent to it.
\item 
The form $\Psi^*\Theta_\Lag$ is closed.
\end{enumerate}
\end{teor}

In coordinates, $\omega = W^i \d^{m-1}x_i$, and the
Hamilton--Jacobi equation in the Lagrangian formalism has the form
$$
\sum_{i=1}^{m}\derpar{W^i}{x^i} + \psi_i^\alpha\derpar{W^i}{u^\alpha} - L(x^i,u^\alpha,\psi^\alpha_i) = 0 \, ,
$$

\begin{definition}
Let $\Lambda\subseteq\Real^{mn}$.
A family of  solutions$\{\Psi_\lambda; \lambda\in\Lambda\}$,
depending on $n$ parameters
$\lambda\equiv (\lambda_1,\ldots,\lambda_n)\in\Lambda$,  is a  
\textbf{complete solution to the Lagrangian Hamilton--Jacobi problem} 
if the map 
$$
\begin{array}{ccccc}
\Phi & \colon &E\times\Lambda & \longrightarrow & J^1\pi \\
 & & (p,\lambda) & \mapsto & \Psi_\lambda(p)
\end{array}
$$
is a local diffeomorphism.
\end{definition}

A complete solution defines an $(m-n)$-dimensional foliation in $J^1\pi$ which is transverse to the fibers
and such that the distribution ${\cal D}_\L$ is tangent to it.
Then, all the sections which are solutions to the Euler--Lagrange equations 
(that is, all the integral sections of the distribution ${\cal D}_\L$)
are recovered from a complete solution.

\subsubsection{Multisymplectic Hamiltonian Hamilton--Jacobi problem}

The Hamiltonian formalism for a regular first-order multisymplectic field theory
is developed in the so-called {\sl reduced dual jet bundle of $J^1\pi$},
$J^1\pi^*= \Lambda^{m}_{2}(\Tan^*E)/\Lambda^{m}_{1}(\Tan^*E)$
(where $\Lambda^m_2(\Tan^*E)$ is the bundle of  
$m$-forms over $E$ vanishing when they act on $\pi$-vertical bivectors).
It is endowed with the canonical projections $\pi_E \colon J^{1}\pi^* \to E$ and
$\bar{\pi}_E \colon J^{1}\pi^* \to M$,
and natural coordinates in $J^1\pi^*$ are denoted $(x^i,y^\alpha,p_\alpha^i)$.
The physical information is given by a {\sl Hamiltonian section} $h$ of the 
natural projection 
$\mu \colon \Lambda^{m}_{2}(\Tan^*E) \to J^{1}\pi^*$,
which is associated with a {\sl local Hamiltonian function} $H \in C^\infty(J^1\pi^*)$ 
such that $h(x^i,y^\alpha,p_\alpha^i) = (x^i,y^\alpha,-H,p_\alpha^i)$. 
Then, from the canonical
form $\Omega \in\df^{m+1}(\Lambda^m_2(\Tan^*E))$, 
we construct the Hamilton-Cartan multisymplectic form
$\Omega_h = h^*\Omega\in \df^{m+1}(J^1\pi^*)$ whose coordinate expression is
$$
\Omega_h = -\d p_\alpha^i \wedge \d y^\alpha \wedge \d^{m-1}x_i + \d H \wedge \d^{m}x \ ,
$$
and the couple $(J^1\pi^*,\Omega_h)$ is a {\sl multisymplectic Hamiltonian system}. 
Then, the {\sl Hamiltonian problem} consists in finding integrable $m$-dimensional
$\bar{\pi}_E$-transverse distributions ${\cal D}_h$ in $J^1\pi^*$ such that their integral sections
$\psi_h \in \Gamma(\bar{\pi}_E)$ are solutions to the
Hamiltonian field equation
$$
\psi_h^*\inn(X)\Omega_h = 0 \, , \ \mbox{\textit{for every} } X \in \vf(J^1\pi^*) \ .
$$
The existence of such distributions ${\cal D}_h$ is assured.
In coordinates, this equation gives the
{\sl Hamilton--De Donder--Weyl equations}
$$
 \derpar{(y^A\circ\psi_h)}{x^\nu}=
 \derpar{{\rm h}}{p^\nu_A}\circ\psi_h
\quad ,\quad
 \derpar{(p_A^\nu\circ\psi_h)}{x^\nu}=
 - \derpar{{\rm h}}{y^A}\circ\psi_h \ .
$$

\begin{definition}
\label{multigenHJh}
The \textbf{generalized Hamiltonian Hamilton--Jacobi problem} 
for the multisymplectic Hamiltonian system $(J^1\pi^*,\Omega_h)$  
is to find a section $s \in \Gamma(\pi_E)$ 
and an integrable $m$-dimensional distribution ${\cal D}$ in $E$ such that,
if $\gamma \in \Gamma(\pi)$ is an integral section of ${\cal D}$, then
$\psi_h=s \circ \gamma \in \Gamma(\bar{\pi}_E)$ is an integral section of ${\cal D}_h$; that is, if
$\Tan_{u}{\rm Im}(\gamma) = {\cal D}_{u}$, for every $u\in{\rm Im}(\gamma)$,
then $\Tan_{\bar u}{\rm Im}(s \circ \gamma) = ({\cal D}_h)_{\bar u}$,
for every $\bar u \in {\rm Im}(s \circ \gamma)$.
Then, the couple $(s,{\cal D})$ is a \textbf{solution to the generalized 
Hamiltonian Hamilton--Jacobi problem}.
\end{definition}

\begin{remark}{\rm
As in the Lagrangian case, the Hamiltonian Hamilton--Jacobi problem
can be considered as a special case of the following {\sl slicing problem}
$$
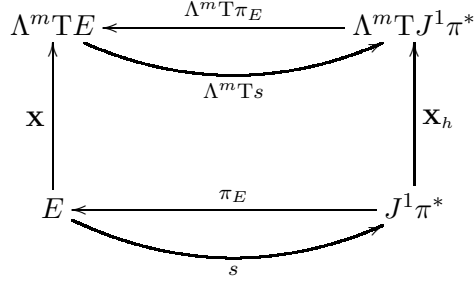
\xymatrix{
\Lambda^m\Tan E \ar@/_1.5pc/[rrr]_{\Lambda^m\Tan s} & \ & \ &
 \ar[lll]_-{\Lambda^m\Tan\pi_E} \Lambda^m\Tan J^1\pi^* \\
\ & \ & \ & \ \\
E \ar[uu]^{{\bf X}} \ar@/_1.5pc/[rrr]_{s} & \ & \ &
 \ar[lll]_{\pi_E}J^1\pi^* \ar[uu]_{{\bf X}_h}
}
$$
where ${\bf X}$ and ${\bf X}_h$ are $m$-multivector fields
associated with the distributions ${\cal D}$ and  ${\cal D}_h$, respectively.
}
\end{remark}

The following Theorems and Definitions are analogous to those of the Lagrangian case.

\begin{teor}
\label{prop:HamGenHJEquivalences}
The following conditions are equivalent.
\begin{enumerate}
\item
The couple $(s,{\cal D})$ is a solution to the generalized 
Hamiltonian Hamilton--Jacobi problem.
\item
The distributions ${\cal D}$ and ${\cal D}_h$ are $s$-related.
As a consequence, the distribution ${\cal D}$ is given by
${\cal D} = \Tan\pi_E({\cal D}_h\vert_{{\rm Im}(s)})$, and it is called the
{\rm distribution associated with $s$}.
\item 
The distribution ${\cal D}_h$ is tangent to the submanifold ${\rm Im}(s)$
of $J^1\pi^*$.
\item
The integral sections of ${\cal D}_h$ which have boundary conditions in ${\rm Im}(s)$ 
project onto the integral sections of ${\cal D}$.
\item 
If $\gamma$ is an integral section of the distribution
${\cal D}$ associated with $s$ then, for every $Y \in \vf(E)$, the equation 
$\gamma^*\inn(Y)\d(h \circ s) = 0$
holds for $s$.
\end{enumerate}
\end{teor}

\begin{definition}\label{def:HamHJDef}
The \textbf{Hamiltonian Hamilton--Jacobi problem} 
for the multisymplectic Hamiltonian system $(J^1\pi^*,\Omega_h)$ 
is to find a section
$s \in \Gamma(\pi_E)$ such that it is a solution to the generalized Hamilton--Jacobi problem
and satisfies that $s^*\Omega_h = 0$.
Then, the section $s$ is a \textbf{solution to the Hamiltonianian Hamilton--Jacobi problem}.
\end{definition}

\begin{teor}
\label{prop:HamHJEquivalences}
The following conditions are equivalent.
\begin{enumerate}
\item
The couple $(s,{\cal D})$ is a solution to the generalized 
Hamiltonian Hamilton--Jacobi problem.
\item 
${\rm Im}(s)$ is an $m$-Lagrangian submanifold of $J^1\pi^*$ and the distribution ${\cal D}_h$ is tangent to it.
\item 
The form $h \circ s \in \df^{m}(E)$ is closed.
\end{enumerate}
\end{teor}

As the $\pi_E$-semibasic $m$-form $h \circ s$ is closed,
there exists a local $\pi$-semibasic $(m-1)$-form $\omega \in \df^{m-1}(E)$, such that
$h \circ s=\d\omega$. In coordinates, if $\omega = W^i \d^{m-1}x_i$, where $W^i \in C^\infty(E)$
are local functions, we obtain that
$$
-H(x^i,y^\alpha,s_\alpha^i) = \sum_{i=1}^{m} \derpar{W^i}{x^i} \quad ; \quad
\derpar{W^i}{y^\alpha} = s_\alpha^i \, ,
$$
from where we obtain the classical Hamiltonian Hamilton--Jacobi equation
$$
\sum_{i=1}^{m} \derpar{W^i}{x^i} + H\left(x^i,y^\alpha,\derpar{W^i}{y^\alpha}\right) = 0 \, .
$$


The definition and the characteristics of complete solution are similar to those
of the Lagrangian case.

\subsubsection{Relation between the multisymplectic Hamilton--Jacobi problems}

Finally, let ${\cal FL}\colon J^1\pi\to J^1\pi^*$ be the Legendre transformation
defined by the Lagrangian $\L$, which is locally given by
\begin{equation*}
{\cal FL}^*x^i = x^i \quad , \quad 
{\cal FL}^*y^\alpha = y^\alpha \quad , \quad
{\cal FL}^*p_\alpha^i = \derpar{\L}{y_i^\alpha} \, .
\end{equation*}
If $\L$ is a regular or a  hyperregular Lagrangian (i.e., ${\cal FL}$ is
a local or global diffeomorphism),
then ${\cal FL}^*\Theta_h = \Theta_\Lag$ and ${\cal FL}^*\Omega_h = \Omega_\Lag$.
In addition, the integral sections of the distributions
${\cal D}_\Lag$ and ${\cal D}_h$, which are the solution to the Lagrangian and 
the Hamiltonian problems respectively,
are in one-to-one correspondence through ${\cal FL}$. 
(See \cite{LPRV-2017} for definitions and details).
Then we have:

\begin{teor}\label{thm:EquivalenceLagHam}
Let $\Lag \in \df^{m}(J^1\pi)$ be a regular or a hyperregular Lagrangian.
Then, if $\Psi \in \Gamma(\pi^1)$ is a jet field solution to the (generalized) Lagrangian
Hamilton--Jacobi problem, then the section $s={\cal FL}\circ \Psi \in \Gamma(\pi_E)$ is a solution to
the (generalized) Hamiltonian Hamilton--Jacobi problem.
Conversely, if $s \in \Gamma(\pi_E)$ is a solution to the (generalized) Hamiltonian
Hamilton--Jacobi problem, then the jet field 
$\Psi ={\cal FL}^{-1} \circ s \in \Gamma(\pi^1)$ is
a solution to the (generalized) Lagrangian Hamilton--Jacobi problem.
\end{teor}
\proof
(Guidelines for the proof):
The proof follows the same patterns as Theorem \ref{relation},
but using multivector fields.
\qed

\begin{remark}{\rm
As a final remark, notice that the Hamilton--Jacobi theory for {\sl non-autonomous} 
(i.e., {\sl time-dependent\/}) {\sl dynamical systems}
can be recovered from the multisymplectic Hamilton--Jacobi theory
as a particular case taking $M=\Real$ and identifying
the distributions ${\cal D}$, ${\cal D}_\L$, ${\cal D}_h$ and their
associated multivector fields ${\bf X}$, ${\bf X}_\L$, ${\bf X}_h$
with time-dependent vector fields (see \cite{LPRV-2017}).
}\end{remark}

\section{Conclusions and outlook}
\protect\label{conc}

In this work, the Lagrangian and the Hamiltonian versions of the
Hamilton--Jacobi theory has been reviewed from a modern geometric perspective.

First, this formulation is done for autonomous dynamical systems
and, in particular, the Hamiltonian case is compared with the 
 ``classical'' Hamiltonian Hamilton--Jacobi theory which 
is based in using canonical transformations.

There is also a general framework for the theory, 
which is also reviewed in the work.
It contains the above standard theory
for autonomous dynamical systems as a particular case, and
allows us to extend the Hamilton--Jacobi theory to a 
wide range of physical systems.
In particular two of these extensions have been analyzed here:
the higher-order (autonomous) dynamical systems 
and the (first-order) classical Lagrangian and Hamiltonian field theories,
using their multisymplectic formulation.

This geometric model has been extended and applied
to many kinds of physical systems (as it is mentioned in the Introduction
and cited in the bibliography).
As a future line of research that has not been explored yet,
the application of this geometric framework to state
the Hamilton--Jacoby equation for dissipative systems in classical field theories
should be explored, using an extension of the
contact formalism which has been recently introduced to
describe geometrically these kinds of dissipative field theories
\cite{GGMRR-2019,GGMRR-2020}.


\section*{Acknowledgments}

I acknowledge the financial support from
project PGC2018-098265-B-C33 of the Spanish
Ministerio de Ciencia, Innovaci\'on y Universidades 
and the project 2017--SGR--932 of the Secretary of University and Research of the Ministry of Business and Knowledge of the Catalan Government




\end{document}